\documentclass[fleqn,usenatbib]{mnras}

\usepackage{newtxtext,newtxmath}

\usepackage[T1]{fontenc}

\DeclareRobustCommand{\VAN}[3]{#2}
\let\VANthebibliography\thebibliography
\def\thebibliography{\DeclareRobustCommand{\VAN}[3]{##3}\VANthebibliography}

\usepackage{graphicx}	
\usepackage{amsmath}	

\usepackage{mwe,tikz}\usepackage[percent]{overpic}
\usepackage{subcaption}

\title[Short title, max. 45 characters]{Image plane detection of FRB121102 with the MeerKAT radio telescope}

\author[Andrianjafy et al.]{
J. C. Andrianjafy,$^{1}$\thanks{E-mail: andrianjafyjuliocsar@yahoo.fr}
N. Heeralall-Issur,$^1$ 
A. A. Deshpande$,^{2,3,4}$ 
K. Golap,$^5$
P. Woudt,$^6$
M. Caleb,$^{7,8}$
\newauthor
E. D. Barr,$^9$
W. Chen,$^9$
F. Jankowski,$^{10}$
M. Kramer,$^{9,10}$
B. W. Stappers,$^{10}$
and J. Wu$^9$
\\
$^{1}$Department of Physics, University of Mauritius, R\'eduit 80837, Mauritius\\
$^2$Raman Research Institute, Bangalore 560080 , India\\
$^3$Inter-University Centre for Astronomy and Astrophysics, Pune 411007, India\\
$^4$Indian Institute of Technology, Kanpur 208016, India\\
$^5$National Radio Astronomy Observatory, PO Box O, Socorro, NM 87801, USA\\
$^6$Department of Astronomy, University of Cape Town, Private Bag X3, Rondebosch, 7701, South Africa\\
$^7$Sydney Institute for Astronomy, School of Physics, The University of Sydney, Sydney, 2006, NSW, Australia\\
$^8$ASTRO3D: ARC Centre of Excellence for All-sky Astrophysics in 3D, Canberra, 2601, ACT, Australia\\
$^9$Max-Planck-Institut f{\"u}r Radioastronomie, Auf dem H{\"u}gel 69, D-53121 Bonn, Germany\\
$^{10}$Jodrell Bank Centre for Astrophysics, Department of Physics and Astronomy, The University of Manchester, Manchester M13 9PL, UK\\
}

\date{Accepted XXX. Received YYY; in original form ZZZ}

\pubyear{2022}

\begin{document}
\label{firstpage}
\pagerange{\pageref{firstpage}--\pageref{lastpage}}
\maketitle

\begin{abstract}
We present the analysis of radio interferometric 2-s images from a MeerKAT observation of the repeating fast radio burst FRB121102 on September 2019, during which 11 distinct pulses have been previously detected using high time and frequency resolution data cubes. In this work, we detected 6 out of the 11 bursts in the image plane at 1.48 GHz with a minimum peak signal-to-noise ratio (S/N) of 5 $\sigma$ and a fluence detection limit of $\sim $ 0.512 Jy ms. These constitute the first detections of a fast radio burst (FRB) or a radio transient using 2-s timescale images with MeerKAT data. Analysis of the fitted burst properties revealed a weighted average precision of $\sim$ 1 arcsec in the localization of the bursts. The accurate knowledge of FRB positions is essential for identifying their host galaxy and understanding their mysterious nature which is still unresolved to this day. We also produced 2-s images at 1.09 GHz but yielded no detection which we attributed to the spectral structure of the pulses that are mostly higher in strength in the upper frequencies. We also explore a new approach to difference imaging analysis (DIA) to search for transients and find that our technique has the potential to reduce the number of candidates and could be used to automate the detection of FRBs in the image plane for future MeerKAT observations. 
\end{abstract}




\section{Introduction}\label{sec:Introduction}
Fast radio bursts  (FRBs) are the newly discovered bright $\sim$ 1 Jy $\sim$ millisecond duration  radio transients \citep{lorimer2007bright}. The source of FRBs emission is yet of unknown origin but the foremost leading theory suggests magnetars as their progenitors \citep{li2021hxmt,scholz2020bright,bochenek2020fast,mereghetti2020integral}. The mysterious nature of FRBs aroused tremendous consideration from the astronomy community and there have been several advancements made during the past decade to understand these phenomenon \citep[for reviews, see ][]{caleb2021decade,petroff2019fast}. 

Given their high range of dispersion measure values, FRBs are potentially invaluable tools to probe the cosmological unverse such as the study of intergalactic turbulence \citep{zhu2021dispersion}, the determination of the cosmic baryon density in the intergalactic medium \citep{macquart2020census}, as well as an estimation of the Hubble constant, being suggested by \citet{wei2019constraining}. However, these studies can only be operated with an accurate knowledge of the position of the FRB source and its host galaxy.    

One of the major findings in FRB field is the first detection of the repeating source FRB121102 \citep{spitler2016repeating,scholz2016repeating} with the single dish Arecibo telescope. The repetition of the bursts allowed targeted follow-up observations with radio interferometers to measure its position up to $\sim$ milliarcsecond precision using high resolution fast timescale imaging \citep[with coordinates  $\alpha = 05^\textrm{h}31^\textrm{m}58.698^\textrm{s}, \delta = 33^{\circ}08'52.586''$;][]{marcote2017repeating, chatterjee2017direct}. As a result, \citet{tendulkar2017host} found that FRB121102 is localized in a low-mass and low-metallicity dwarf galaxy by matching the measured position with optical observations from the Gemini North telescope. Among the few hundreds distinct FRBs reported  in the literature \citep{amiri2021first,petroff2016frbcat}, more than 10 of them have been localized to their host galaxies by combining the interferometric image plane location of the bursts and its matched position with telescopes operating at other wavelengths. For instance, \citet{ravi2019fast} localized FRB190523 with the use of $\sim$0.5-s radio images from the Deep Synoptic Array (DSA) and the low resolution imaging spectrometer of the Keck I telescope. Similarly, the position of FRB190711 were obtained with 3.1-s radio images from the Australian Square Kilometre Array Pathfinder (ASKAP) and deep images of the Very Large Telescope \citep{macquart2020census}.  
  
In general, detecting FRB and fast transient type sources through interferometric imaging can be achieved by correlating the signals from each pair of antennas in the array and averaging the correlated output, known as \textit{visibilities}, in short integration time that is adequate to the science requirements. However, searching for fast radio transients in the image plane can encounter different challenges. The sparse $\textit{(u,v)}$ sample over the short period of time could give rise to lower sensitivity and poor quality images. The duration of the transient pulses is often much shorter than the integration time, causing loss of signals into the noisy visibility data. The other challenging factor is also the high rate at which data is recorded that makes computing heavily expensive, and often preventing real time imaging detection.

In 2019, observations of the repeating source FRB121102 have been performed with the MeerKAT radio telescope in South Africa. From this observation, 11 bursts from FRB121102 have been detected using high time and frequency resolution data cubes \citep{caleb2020simultaneous} by the The Meer(more) TRAnsients and Pulsars (MeerTRAP) team \citep{stappers2016meertrap}. Knowing the position and arrival times of these pulses, these detections are of crucial importance in investigating and testing the capability of the MeerKAT telescope to detect FRBs in the image plane using fast dump visibility data, which will be the main focus of this paper.  

The paper is organized as follows, the observations are described in Section \ref{sec:dataobservations}, followed by the description of our methodology including data flagging, calibration and imaging in Section \ref{subsec:editing} and \ref{subsec:calibration}. The results are presented in Section \ref{sec:results} and their properties will be discussed in Section \ref{sec:discussion} as well as its implications towards MeerKAT surveys before we conclude in Section \ref{sec:conclusion}.  

\section{Methods} \label{sec:methods} 
\subsection{Data observations} \label{sec:dataobservations} 
As part of a Director’s Discretionary Time (DDT) proposal, MeerKAT \citep{mauch20201,jonas2016meerkat} carried out 3 hour observations towards the position of FRB121102 on the 6th September 2019, 10th September 2019, 6th October 2019, and 8th October 2019. The 4 sessions were conducted in slightly different telescope configurations (number of antennas, integration time, frequency channel resolution), therefore we will only describe the 10th September observation during which FRB121102 pulses have been previously detected and for which we performed our analysis. In our data, the observations were conducted using 58 out of the 64 MeerKAT dishes at a frequency center of 1.28 GHz and a total bandwidth of 856 MHz that is divided into 4096 frequency channels. Although the data was initially recorded with a native resolution of 4.785 $\mu$s, the visibility data was dumped at 2 s integration time for the imaging purposes. The observations of the target were separated into 12 scans of 15 minutes each. A primary calibrator, J0408-6545, was observed for 5 minutes at the beginning of the observation, as well as a complex gain calibrator J0534+1927, which was observed in two sessions of 1 minute duration before and after the 3 hour observation of the target. To accelerate the data processing in this work (see Section \ref{subsec:editing} and \ref{subsec:calibration}), we only carried out our analysis on a set of 5 minutes data around the reported arrival times of the 11 bursts \citep{caleb2020simultaneous}.

\subsection{Data editing}\label{subsec:editing}
Firstly, we corrected for the observed shift in the time stamps of the visibility data which was offset by 2 s to their true values. This offset error has been fixed for the latest MeerKAT raw data as discussed in \citet{mauch20201}.       

We have decided to unflag all the data flagged by the MeerKAT online flagger \footnote{\url{https://skaafrica.atlassian.net/wiki/spaces/ESDKB/pages/305332225/Radio+Frequency+Interference+RFI}}. This procedure was applied because the effect of the online flagger to transient detection is not well studied and transient emissions could be mistaken as radio frequency interference (RFI). Due to the weak response of the receiver at the edges \citep{mauch20201}, we trimmed 210 and 186 frequency channels at the lower and higher edges of the frequency band. These values were chosen based on manual inspection of the visibility data. Our final bandwith for processing is then ranging from 900 to 1673 MHz. 

To mitigate RFI, frequencies that are known to be corrupted for MeerKAT L-band were flagged for short baselines ($<$1km). Given the large amount of data, automated flagging algorithms are required to remove residual RFI. To this end, we adopted different strategies to flag the calibrators and target data. In the case of calibrators, we run a combination of automated algorithms including, \textit{sumthreshold} in \textup{AOFlagger} \citep[version 3.0.0;][]{offringa-2012-morph-rfi-algorithm}, \textit{tfcrop} and \textit{rflag} from the Common Astronomy Software Applications, \textup{CASA}\citep[version 6.0.0;][]{mcmullin2007casa}. The aggresive flagging applied to the calibrators is necessary to obtain good calibration solutions. 
However, for the target data, we decided to avoid algorithms that flag data based on thresholding in the time direction in order to minimize removal of potential transient candidates that only appear for a short period of time \citep{cendes2018rfi}. We instead chose to use the \textit{threshold\textunderscore channel\textunderscore rms} function in \textup{AOFlagger} which averages data of a given segment length in time and flag frequency channels in which the rms of their amplitude values exceed a user based threshold (3$\sigma$ level in our case). This step was repeated three times during which we vary the segment length in a decreasing time intervals (5 minutes, 1 minute and 10 seconds) to identify channels containing both steady and intermittent RFI. After these procedures, about 40 $\%$ of our target data was removed.

\subsection{Data calibration and imaging}\label{subsec:calibration}
\subsubsection{Initial calibration}\label{subsub:1GC}

Following data flagging, we performed standard data calibration method using \textup{CASA} tools. With a known absolute flux density ($\sim$ 17 Jy at 1.28 GHz and a spectral index of -1.179\footnote{\url{https://github.com/IanHeywood/oxkat/blob/master/oxkat/1GC_04_casa_setjy.py}}), we used the primary calibrator to derive the frequency-dependent complex gain factors. We then bootstrapped its flux density to the secondary calibrator, from which time-dependent gain solutions were determined. The solutions were then applied to all datasets. 

Afterwards, we divided the data into 7 subbands and performed 2 rounds of phase-only self calibration in each of the subbands independently. The sky-models to perform self calibration was generated with the fast generic widefield imager  \textup{WSCLEAN} \citep[version 2.9.0;][]{offringa2017optimized,offringa2014wsclean}. The sky-models from each 5 minute dataset contain sources with high signal-to-noise ratio (S/N $>450$) that is sufficient for our self calibration process to succeed.  

The observations did not contain a polarized calibrator to perform an accurate polarimetric calibration. The other alternative is to use the averaged polarization calibration of the antennas obtained from other MeerKAT data, which was calibrated using a polarized calibrator and apply them to our measurement sets. The latter method, as described in \citet{plavin2020}, was not feasible in our case because it requires the two datasets to have the same reference antenna and there were no available MeerKAT datasets that had polarization calibration using our preferred reference antenna ($''m007''$), which was chosen based on the S/N of the parallel-hand calibration solutions.

\subsubsection{Peeling}\label{subsub:peeling}
We identified two bright sources ($>100$ mJy/beam) dominating the FRB121102 field, that exhibit direction-dependent effects towards the center, especially with the 2-s images. Such effects are undesirable as residual undeconvolved sidelobes from those sources can potentially mimic transients in the image plane \citep{frail2012revised,bower2007submillijansky}.         

Hence, we chose to remove these sources with the technique known as \textit{peeling} \citep{noordam2004lofar}. There are many existing approaches \citep{kazemi2011radio,smirnov2011revisiting,intema2009ionospheric} to perform peeling but we adopted similar strategy to the one described in \citet{williams2019tool}. In this method, the calibration terms $G_{bs}$ specific to the bright source to be peeled is determined and it can be subtracted by replacing the model column with:
 
\begin{equation}
MC_{cor} = \frac{1}{G_{bs}}M_{bs}(u,v),
\end{equation}  

 where $MC_{cor}$ is the corrected model column and  $M_{bs}(u,v)$ an approximated model of the bright source in $\textit{(u,v)}$ space.

\begin{figure*}
  \centering   
  \begin{overpic}[scale=0.49]{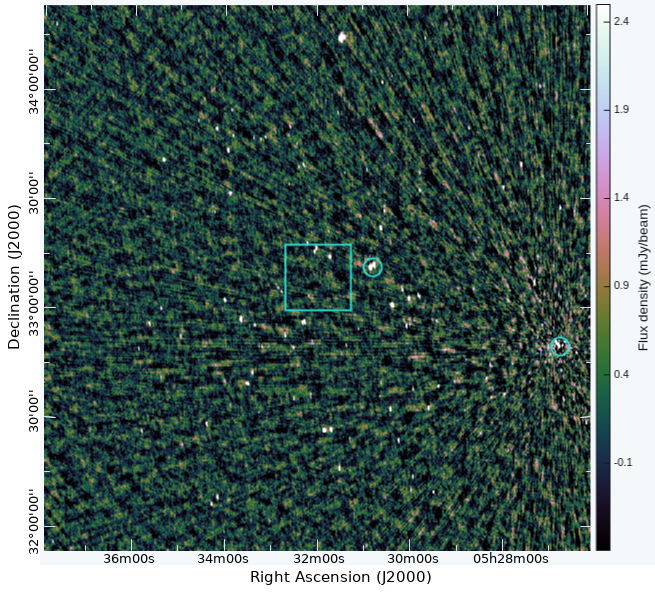}
       \put(9,12){\includegraphics[scale=0.15]{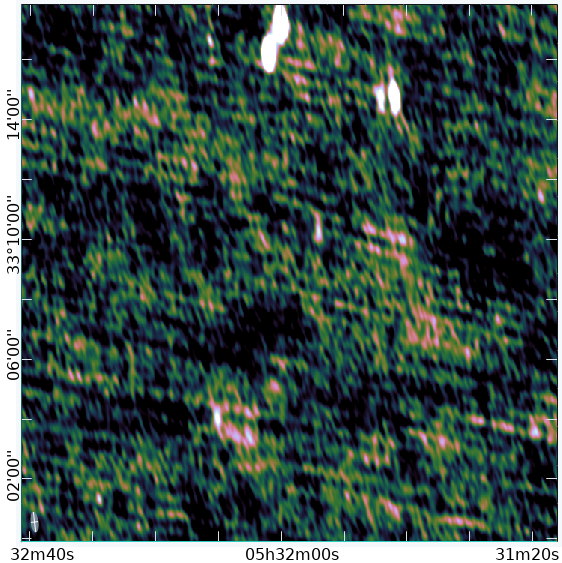}} 
  \end{overpic}
  \begin{overpic}[scale=0.49]{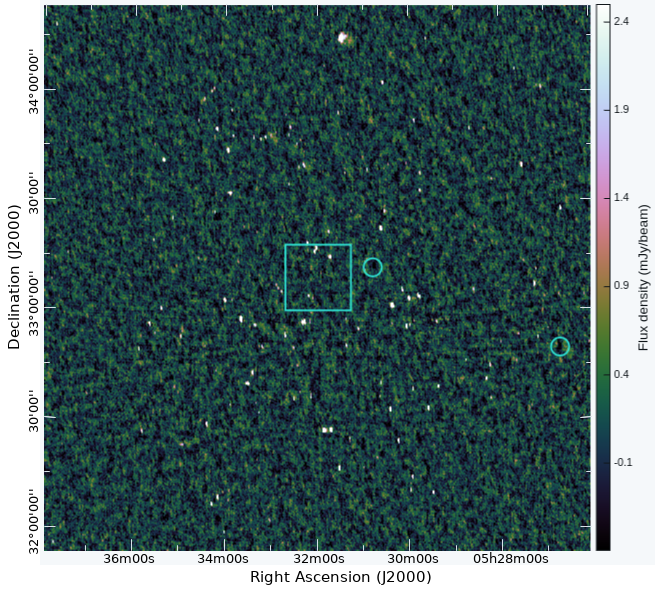}
       \put(9,12){\includegraphics[scale=0.15]{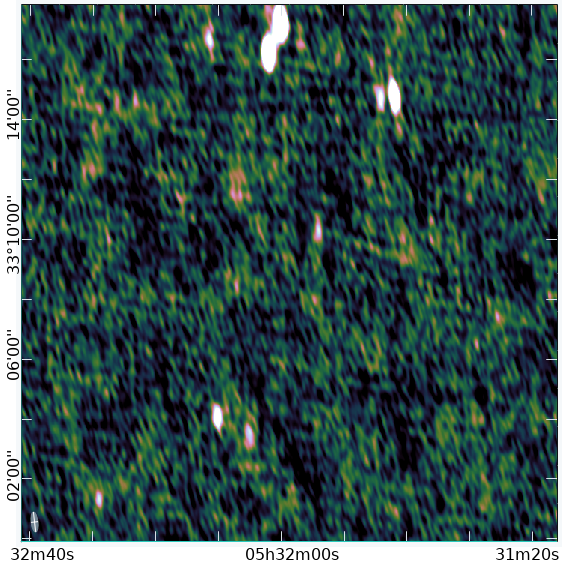}} 
  \end{overpic}
\caption{Comparison of MeerKAT 2-s images of the FRB121102 field at 1.09 GHz before (top panel) and after (bottom panel) peeling two bright sources (blue circles) using the method described in section \ref{subsub:peeling}. The blue squares indicate an area of $18' \times 18'$ around the position of FRB121102.  A zoom in of this region is shown in the bottom left corner of each panel. The improvement in the image quality is clearly seen as the ripples caused by the bright sources was mitigated after peeling, along with the appearance of faint emission from real sources.} 
\label{5min_peel}
\end{figure*}

 In our case, we performed a phase-only followed by a bandpass self calibration towards the source and the inverse of the gains $G_{bs}$ was calculated by dividing the data with the corrected column. $M_{bs}(u,v)$ was generated by predicting the model image of the bright source during self calibration. The bright sources were peeled independently in each 7 subbands to capture the variations of $G_{bs}$ across the frequency band. We implemented these procedures using \textsc{CASA} and \textsc{WSCLEAN} tasks wrapped into customized python scripts. Figure \ref{5min_peel} shows a comparison of the 2-s images at 1.09 GHz (see Section \ref{subsub:imaging} for imaging method) before and after peeling where the rms value at the position of the target went from $533$ to $410$ $\mu$Jy/beam. 
 
 \subsubsection{Imaging}\label{subsub:imaging}
 
 \begin{table}
  \centering
  \caption{\textsc{WSCLEAN} main parameters during 2-s imaging. The other parameters were set to their default values.}
  \begin{tabular}{@{}ccc@{}}
    \hline
    Parameter  & Values \\
    \hline
    size & 6000 \\
    scale & 1.5 arcsec \\
    auto-mask & 4 \\
    auto-threshold & 0.1 \\
    weight & Briggs 1 \\
    super-weight & 10 \\ 
    minuv-l & 100 \\
    taper-inner-tukey & 400 \\ 
    mgain & 0.85  \\
    \hline
  \end{tabular}
\label{imaging_param}
\end{table}

\begin{figure*}
  \centering   
  \begin{overpic}[scale=0.49]{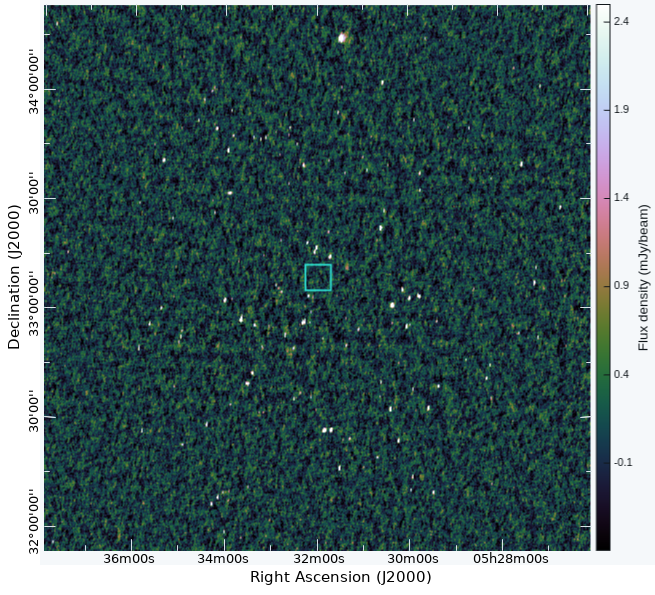}
     \put(58,10){\includegraphics[scale=0.21]{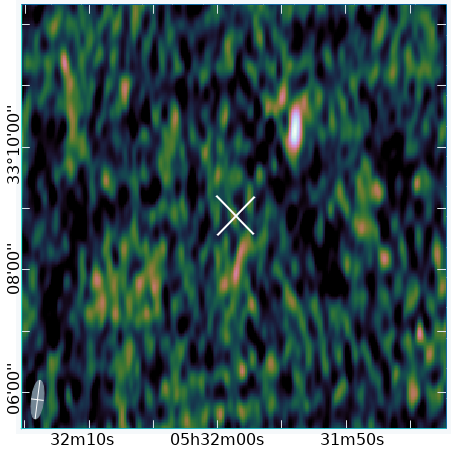}}  
  \end{overpic}
  \begin{overpic}[scale=0.49]{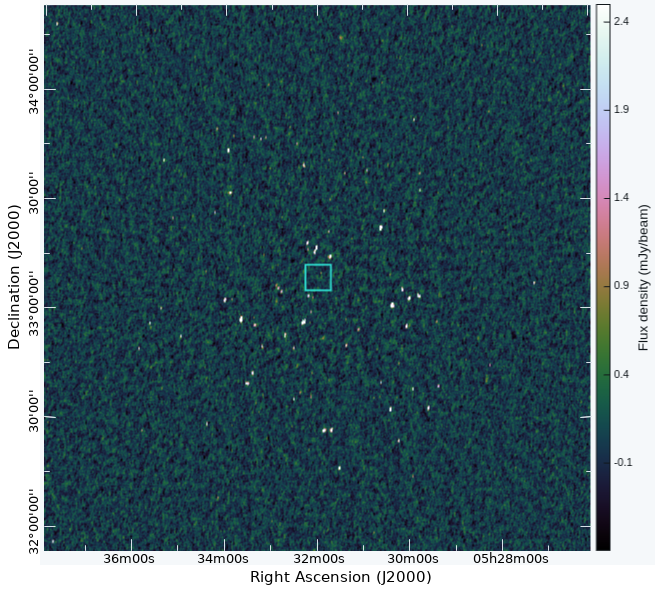}
     \put(58,10){\includegraphics[scale=0.21]{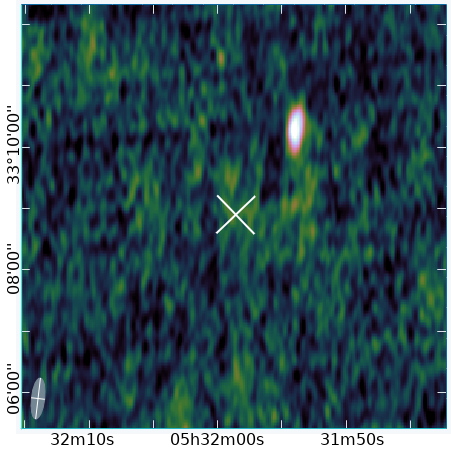}} 
  \end{overpic}
\caption{MeerKAT 2-s images of FRB121102 field at 1.09 GHz (Top) and 1.48 GHz (Bottom) when the bursts are off. The blue squares indicate a $7' \times 7'$ area centered around the position of the FRB and the corresponding zoomed-in view is shown in the bottom right corner of each image. For each zoom-in image, the grey ellipse in the bottom left corner indicates the synthesized beam size and the white cross in the center indicates the reported position of FRB121102. A known $\sim$ 3 mJy compact source, J053153+3310 ($\alpha = 05^\textrm{h}31^\textrm{m}53.92^\textrm{s}, \delta = 33^{\circ}10'20.07''$), is also observed  which we used for astrometry in Section \ref{subsub:astrometry} and Section \ref{subsec:absolute_astrometry}}
\label{2s_images_off}
\end{figure*}

 To search for the bursts, we only divided the data into two frequency bands centered around 1.09 GHz and 1.48 GHz. We review briefly the effect of this spectral window setup in Section \ref{subsec:bandwidth}. We produced stokes I images of each integration for each 5 minutes data set with WSCLEAN using the imaging parameters in Table \ref{imaging_param}. These parameters were tuned to obtain reliable images without decreasing drastically the sensitivity. The automatic masking scheme is suitable to our science goal as it allows deep cleaning close to the thermal noise value, but only constrained towards peaks with high S/N. A circular taper was applied to the inner edges of the $\textit{(u,v)}$ samples as this form of tapering was observed to decrease slightly the level of sidelobes. We did not apply primary beam correction because our target is situated at the phase center. Figure \ref{2s_images_off} illustrates typical examples of the 2-s images of the two subbands when the bursts are not present. 
 
 The rms thermal noise of the images were evaluated within a $7'\times7'$ square around FRB121102 position, using the method described in \citep{swinbank2015lofar} from which pixel values that are more than 4 standard deviations away from the median are masked. Overall, the mean rms is 403 $\mu$Jy/beam and 259 $\mu$Jy/beam at 1.09 and 1.48 GHz with similar elliptically restored beam size of $42''\times13''$. For each sequence of images, we evaluated the peak fluctuation over the same local region and estimated that only pixels with peak S/N $>$5$\sigma$ can be considered as potential transient emission.    

\subsection{FRB limits}\label{subsub:FRB_limit}
Among the main factors deciding the detection of FRB in our image plane is the intrinsic width of the bursts. Considering that the pulse width of the 11 FRBs are much shorter than our integration time, we would expect their peak flux density to decrease due to averaging of the FRB signal in the visibility. Based on the framework described in \citet{trott2013prospects} and \citet{rowlinson2016limits}, the minimum FRB flux density that our 2-s image is sensitive to can be estimated by:
\begin{equation}
S_{min} = S_{snap,2 s} (\frac{\Delta t }{w}), 
\label{eq:frb_limit}
\end{equation} 

where $S_{snap,2 s}$ is the rms in one snapshot 2-s image, $\Delta t$ is 2 s and $w$ the duration of the FRB pulses. From equation (\ref{eq:frb_limit}), we can then define the minimum detectable fluence as:

\begin{equation}
F_{min} = S_{min} w = S_{snap,2 s} \Delta t , 
\label{eq:frb_limit_fluence}
\end{equation}

\begin{figure}
	\centering
	\includegraphics[width=0.47\textwidth, height = 0.47\textwidth]{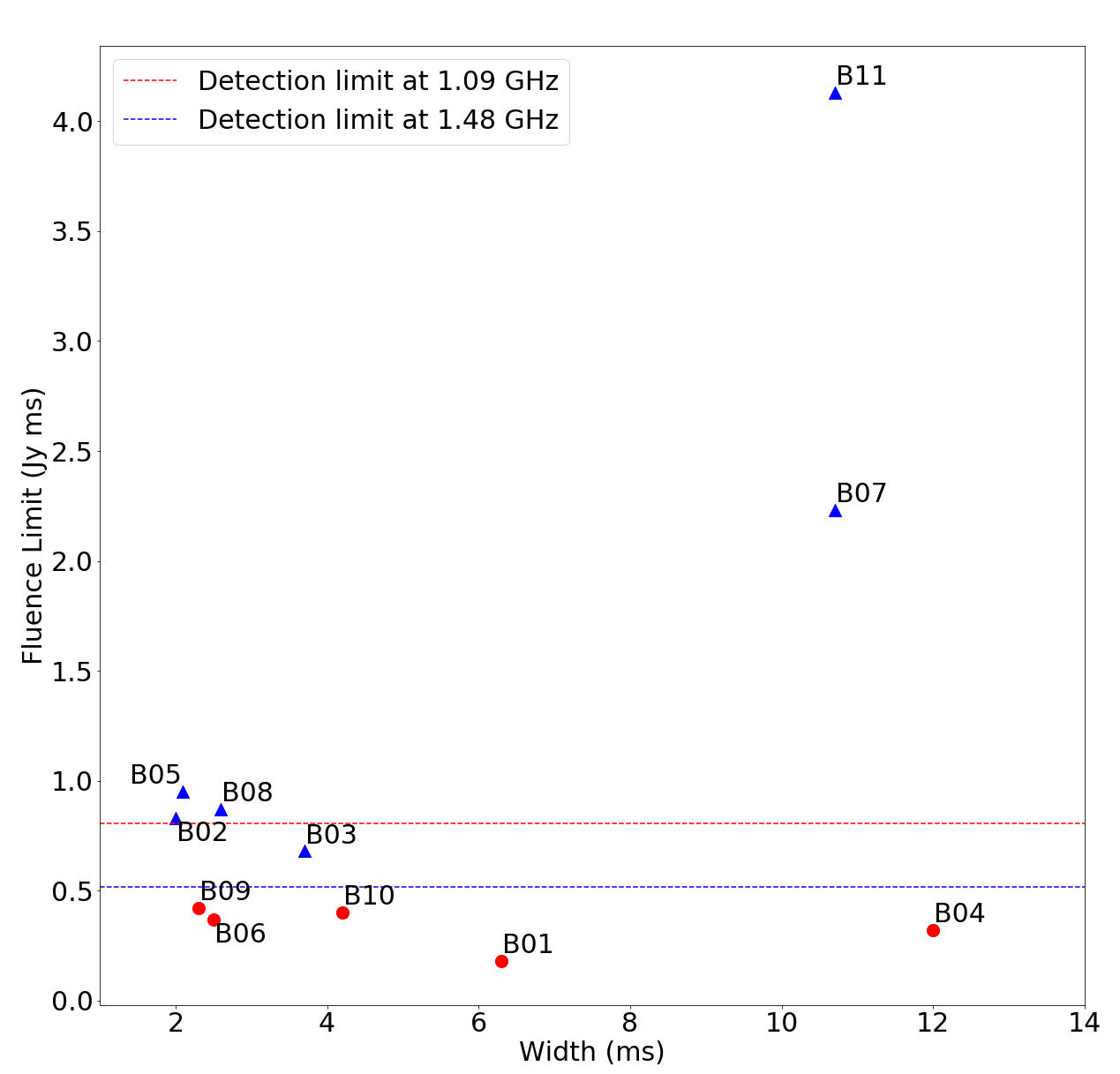}
	\caption{Theoretical fluence limits of detectable FRBs in the MeerKAT 2-s snapshot images produced in this work. The detection limits were calculated using the equation (\ref{eq:frb_limit_fluence}) derived from the frameworks in \citet{trott2013prospects} and \citet{rowlinson2016limits}. The blue triangles are the bursts that we detected at 1.48 GHz while the red points were not observed. These bursts represents all pulses detected in the dynamic spectrum by \citet{caleb2020simultaneous} and referred with the same indices. ie. B01 refers to burst number 01.}
	\label{fig:frb_limit}
\end{figure}

By taking into account the rms values measured in Section \ref{subsub:imaging}, and using the equation (\ref{eq:frb_limit_fluence}), we calculated that the theoretical fluence limit values in our images are 0.80 and 0.51 Jy ms at 1.09 GHz and 1.48 GHz.
Using the fluence values reported in Table 1 of \citet{caleb2020simultaneous}, we could then set constraints on which burst we would expect to detect as illustrated in Figure \ref{fig:frb_limit}, and from which we show that 5 bursts are detectable at 1.09 GHz, while 6 of them lay above the detection limit at 1.48 GHz. 

\section{Results}\label{sec:results}
\subsection{Burst detection} 
\begin{figure*}
	
    \centering
    \begin{subfigure}[b]{0.46\textwidth}
    	\includegraphics[width=\textwidth]{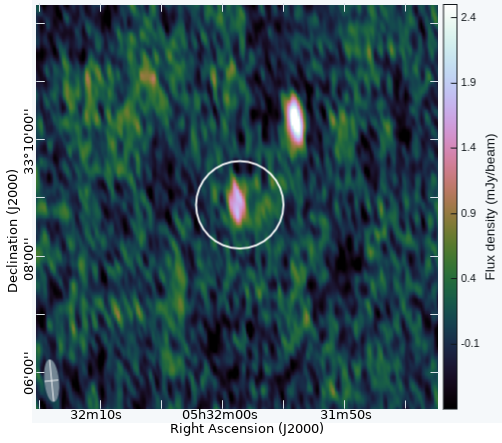}
    	\caption{Burst 02}
    \end{subfigure}
    \centering
    \begin{subfigure}[b]{0.46\textwidth}
    	\includegraphics[width=\textwidth]{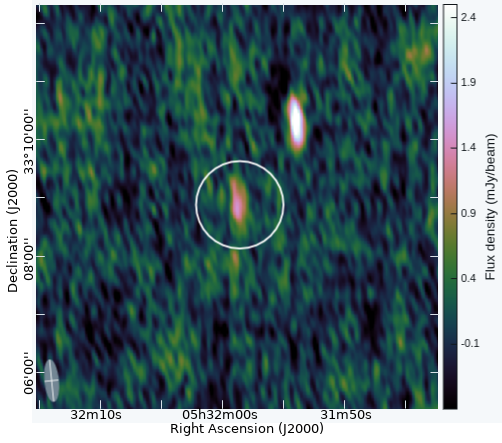}
    	\caption{Burst 03}
    \end{subfigure}
    \centering
    \begin{subfigure}[b]{0.46\textwidth}
    	\includegraphics[width=\textwidth]{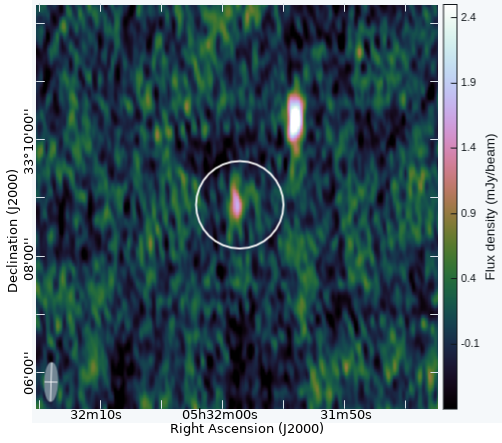}
    	\caption{Burst 05}
    \end{subfigure}
    \centering
    \begin{subfigure}[b]{0.46\textwidth}
    	\includegraphics[width=\textwidth]{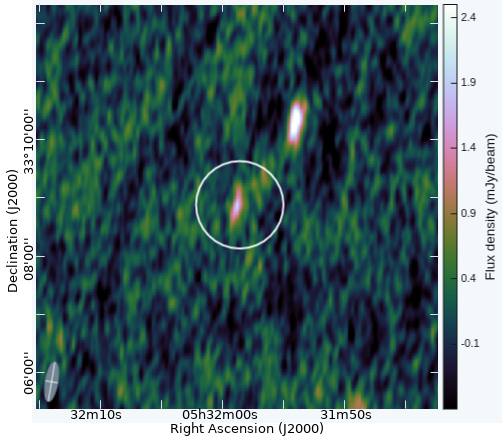}
    	\caption{Burst 07}
    \end{subfigure}

    \centering
    \begin{subfigure}[b]{0.46\textwidth}
    	\includegraphics[width=\textwidth]{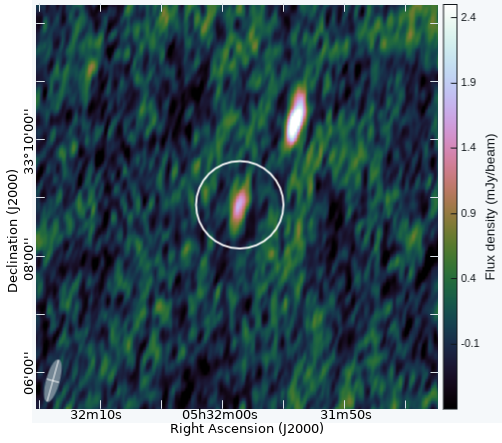}
    	\caption{Burst 08}
    \end{subfigure}
    \centering
    \begin{subfigure}[b]{0.46\textwidth}
    	\includegraphics[width=\textwidth]{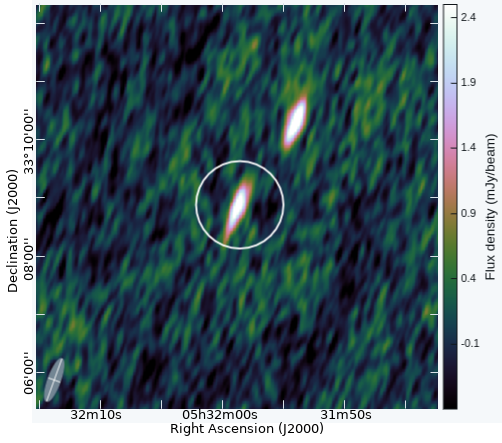}
    	\caption{Burst 11}
    \end{subfigure}
    \caption{MeerKAT 2-s images of the detected bursts (White circles) at 1.48 GHz. The grey ellipse in the bottom left corner indicates the synthesized beam size.}
    \label{Bursts_image}
\end{figure*}

After inspecting the images, we have detected 6 bursts at the arrival times of B02, B03, B05, B07, B08 and B11 (Labelled with the same indices as in \citet{caleb2020simultaneous}) at the position of FRB121102 at 1.48 GHz with a peak S/N above 5$\sigma$. These constitute the first detections of FRBs or any transient sources in 2 seconds radio images with the MeerKAT telescope. Figure \ref{Bursts_image} illustrates the 2-s images during the appearance of all the detected bursts. To estimate the integrated flux density, we fitted a gaussian component with the task \textit{imfit} in \textsc{CASA} in a circular regions that enclose the burst structure. The measured properties of each fitted burst are shown in Table \ref{bursts_properties}. None of these bursts were observed at 1.09 GHz above our peak detection threshold. None of the bursts situated below our detection limits (see Figure \ref{fig:frb_limit}) were detected in the two subbands as expected. 

\subsection{Burst Positions} 
\begin{figure}
	\centering
	
	\includegraphics[width=0.5\textwidth, height = 0.5\textwidth]{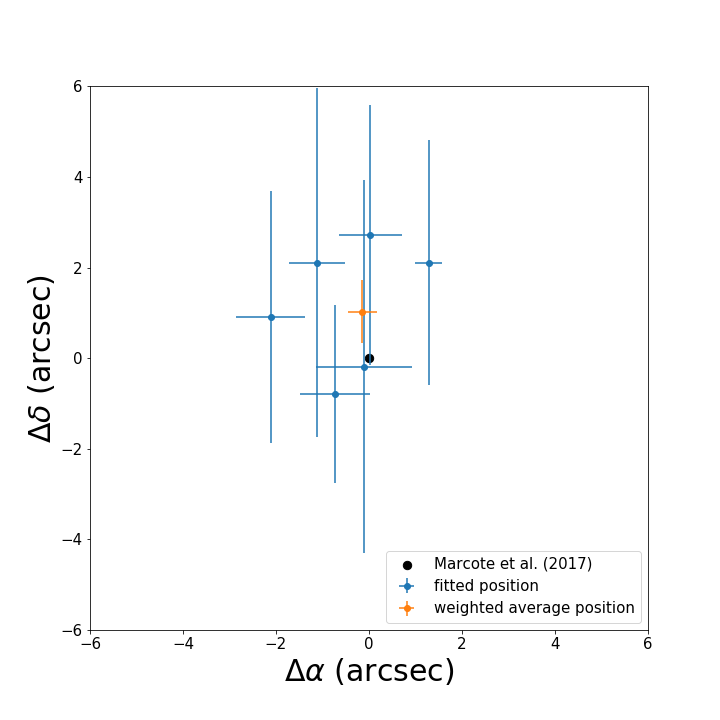}
	
	\caption{Offsets of the fitted centroid position of the FRB121102 bursts detected in this work (Blue points) to the milliarcsecond localization with EVN \citep{marcote2017repeating}. The orange point indicates the weighted average position based on the fitted position uncertainties with CASA. The horizontal lines indicate the position errors for each point.}
	\label{burst_pos_new}
\end{figure}

The fitted centroid location of the bursts are scattered within $\lesssim 2.70''$ of the milliarcsecond localization from the simultaneous observations of European VLBI network and the Arecibo telescopes \citep{marcote2017repeating}. Given that the position uncertainties (see Table \ref{bursts_properties}) are inversely proportional to the source S/N, we estimated that the strongest burst, B11, which is offset by $\sim 1.07''$, gives the highest confidence in our measured positions.    Nevertheless, to account for all the 6 detected bursts,  we calculated the weighted average position of the burst emission peaks by using as weights the inverse of the uncertainties from CASA. As a result, we found a weighted average position of $\alpha = 05^\textrm{h}31^\textrm{m}58.68^\textrm{s} \pm 0.31'', \delta = 33^{\circ}08'53.61''\pm 0.69''$ and a combined offset of $\sim 1.03''$.  The position offsets measured for the individual bursts is shown in Figure \ref{burst_pos_new}.

\begin{table*}
  \centering
  \caption{Properties of the bursts detected in the image plane at 1.48 GHz. The quoted arrival times are the time center of the 2-s visibility data at the moment of detection on 10 September 2019. The peak $P_{\nu}$, integrated flux density $S_{\nu}$ and position values were measured with the \textsc{CASA} task \textit{imfit}. The peak S/N is based on the rms value of the image during detection. The systematic offsets are obtained with the procedure described in Section \ref{subsub:astrometry}}. 
  \begin{tabular}{@{}ccccccccc@{}}
    \hline
    Burst & Arrival time & Peak flux density $P_{\nu}$ & Flux density $S_{\nu}$ & Peak S/N & \multicolumn{2}{|c|}{Centroid position [J2000]} & \multicolumn{2}{|c|}{Systematic offsets}\\  
    
     & (UTC) & (mJy) & (mJy) & & RA (h:m:s) & DEC ($^{\circ}$:$'$:$''$) & RA ($''$) & DEC ($''$) 
    \\
    \hline
    02 & 03:58:31.4 &1.85 $\pm$ 0.21 &  2.22 $\pm 0.46$ & 7.2& 05:31:58.70 $\pm 0.67''$ & 33:08:55.37 $\pm 2.87''$ & $+0.51 \pm 0.36$ & $-0.09 \pm 2.27$\\
    03 & 03:58:33.4 &1.41 $\pm$ 0.15 &  1.93 $\pm 0.14$ & 5.4& 05:31:58.61 $\pm 0.60''$ &33:08:54.75 $\pm 3.86'' $ & $-0.71 \pm 0.26$ & $-1.08 \pm 2.37$\\
    05 & 04:26:10.7 &1.58 $\pm$ 0.15 &  1.32 $\pm 0.30$ & 6.1& 05:31:58.80 $\pm 0.30''$ &33:08:54.76 $\pm 2.70'' $ & $+0.58 \pm 0.14$ & $+2.07 \pm 1.43$\\
    07 & 05:04:37.8 &1.51 $\pm$ 0.22 &  1.66 $\pm 0.48$ & 5.4& 05:31:58.69 $\pm 1.03''$ &33:08:52.49 $\pm 4.12'' $ & $-0.19 \pm 0.42$ & $-1.62 \pm 2.22$\\
    08 & 05:38:38.9 &1.67 $\pm$ 0.17 &  1.60 $\pm 0.48$ & 6.4& 05:31:58.53 $\pm 0.74''$ &33:08:53.54 $\pm 2.78'' $ & $-0.08 \pm 0.49$ & $+0.91 \pm 1.91$\\
    11 & 06:06:04.2 &2.90 $\pm$ 0.16 &  2.83 $\pm 0.35$ & 11.1& 05:31:58.64 $\pm 0.75''$ &33:08:51.81 $\pm 1.96'' $ & $+0.03 \pm 0.58$& $-0.47 \pm 1.62$\\
    \hline
  \end{tabular}
  \label{bursts_properties}
\end{table*}

\section{Discussion}\label{sec:discussion}
\subsection{Bursts S/N}
In comparison to our detection limit in Section \ref{subsub:FRB_limit}, B03 and B11 have the lowest and highest S/N as expected. However, we could notice that the fitted peak S/N for B07 in the 2-s image is among the lowest despite its high fluence value of $\sim$ 2.23 Jy ms (see Figure \ref{eq:frb_limit_fluence}). Similarly, we measured a high S/N of B02 ($>$ 7.2) although it is among the bursts with the lowest fluence. The discrepancies between the observed and expected S/N could arise from the variation of rms between the images. At the arrival time of B07 for example, the rms is slightly higher than the average value ($\sim 274 \;\mu$Jy), inducing \textit{imfit} to decrease the fitted peak to $\sim 1.51 $ mJy although the peak directly extracted from the image pixel values is $1.94 $ mJy. Further discussion of flux density accuracy is discussed in the next section.    

\subsection{Flux density and position uncertainty}\label{subsub:astrometry}

Due to the sparse $\textit{(u,v)}$ coverage of the 2 s data, the measured flux density and position from the gaussian fitting can be affected by systematic errors due to residual calibration effects or by the fluctuations of thermal noise in the images. We use the observed compact source, J053153+3310 (with coordinates  $\alpha = 05^\textrm{h}31^\textrm{m}53.92^\textrm{s}, \delta = 33^{\circ}10'20.07''$ from \citet{marcote2017repeating}), with a flux density range comparable to our bursts ($\sim 3$ mJy, $\sim 11\sigma$), to evaluate the flux density and position uncertainties. Given the short angular separation of the compact source from the target (offset by $\sim 100''$ which is about three synthesized beam widths away), we estimated that both sources are affected to the same variation of noises and share relatively the same systematic uncertainties in their fitted values.

Hence, with the 2-s snapshot images in each 5 minute data, we fitted the compact source with \textit{imfit} the same way as the bursts. We defined its fractional flux density error as the ratio of the root mean square of the flux density variation errors with the mean flux density in each 5 minutes epoch. As a result, we obtained an average fractional error of 12 $\%$ in flux density values. Furthermore, we did not observe any significant increase in flux density greater than the fractional error in the compact source during the appearances of the bursts. However, the S/N of the compact source in the B07 image decreased by 1, which indicates that the fitted properties of B07 could be under estimated.     

We evaluated the systematic offsets in our position by comparing the MeerKAT position of the compact source to the VLBI observations. The magnitude of the offsets has a median of $0.45''$ in RA and $1.28''$ in Dec, with interquartile ranges of $0.57''$ and $1.62$ respectively. The systematic offsets measured at the arrival time of each burst is shown in Table \ref{bursts_properties}.    

\subsection{Astrometry}\label{subsec:absolute_astrometry}
The early MeerKAT data suffered from few instrumental issues that could cause systemic inaccuracies of the astrometry \citep{heywood20221,knowles2022meerkat,mauch20201}. We investigated if the discrepancies in our burst positions were the results of these bugs or due to the limited dynamic range of the 2-s images. To this end, we merged all the 5 minute chunks of data in our analysis and imaged the concatenated measurement set ($\sim$ 50 minute observations) to assess the position of J053153+3310 in a higher dynamic range image. We obtained a noise level of 42 $\mu$Jy, yielding a S/N of $\sim$70 for this compact source. With this deeper image, the fitted position of J053153+3310 now deviates by $0.07'' \pm 0.03$ in RA and $0.46'' \pm 0.13$ in DEC from its catalogue position, which is smaller compared to the median offsets observed in the 2-s images, and suggesting that the large position uncertainties are mainly from statistical origin.   

Further testing were applied based on cross-correlation operation between the 2-s images to probe if the peak of the cross-correlated image exhibit a relative shift from the origin, which would indicate the presence of astrometric errors of all the sources in the field resulting from calibration or related to beam shape differences. Therefore, we chose one reference image at the beginning of the observation and cross-correlate it with the 2-s images prior to all the burst appearances. As a consequence, we did not observe a shift in the output peaks with pre-burst images near (in time) the reference image. However, a 1-pixel shift in the declination axis was always noticed for images separated by more than $\sim1$ hour to the reference image. Given that the point spread function (PSF) major axis in our images is elongated along the declination axis, these analysis suggest an astrometric uncertainty of 1.5'' (size of one pixel) due to beam shape in our DEC position.   

By taking into account the S/N and beam width, we estimated that the spread of the FRBs and their position uncertainties in our 2-s images is fairly comparable to the expectation and astrometry correction is not required for the purpose of the present work. The low fractional error and position uncertainty that we obtained decrease the probability that the bursts that we have detected could be produced from imaging artefacts. 

\subsection{Non detection at 1.09 GHz}
Despite the high range values in the rms of the images at 1.09 GHz ($\sim 400 \mu $Jy), some of the bursts that we have detected are still above the detection limit at this frequency but yielded no detection. The produced residual images at the burst arrival times were verified visually but revealed no significant peaks above the 5$\sigma$ detection threshold. Considering that FRB121102 pulses showed some spectral variations, we suspect that the non detection could be explained by the fact that these bursts peak at the higher frequencies as can be seen with their dynamic spectra in Figure 1 of \citet{caleb2020simultaneous}. Spectral index measurements could further support this explanation but the quality of the images declined rapidly in images produced with shorter frequency resolution, making high level of uncertainty in flux density estimation. 

\subsection{Review methods for future MeerKAT observations}
In the next few subsections, we will briefly review and discuss the methodology that we performed in Section \ref{sec:methods} and discuss their practical implication in searching for FRBs in future MeerKAT surveys or archival data.  
\subsubsection{Flagging}
The flagging approach that we described in Section \ref{subsec:editing} efficiently removed all suspected RFI without negatively affecting the detection of the bursts. Manual inspection of the visibility data at the moment of appearance was still checked and we flagged small residual RFI and the bursts were still observed in the images. To investigate the efficiency of our method, we decided to apply to our data the automated flagging algorithm \textit{Tricolour}\footnote{\url{https://github.com/ratt-ru/tricolour}}, which is widely incorporated into some of the latest MeerKAT data reduction pipelines such as MeerKATHI\footnote{\url{https://meerkathi.readthedocs.io/en/latest/}} \citep{jozsa2020meerkathi} or \textit{oxkat}\footnote{\url{https://ascl.net/code/v/2627}}\citep{heywood2020oxkat}. As a results, the S/N for B05, B08, and B07 decreased by $\sim 10 \%$ and B02 was not detected. Although, we tested \textit{Tricolour} in its default mode and a customized strategy might be required to optimize the algorithm, these tests show how an automated flagging technique using thresholding following the time axis can affect the detection of short transient emissions. 

\subsubsection{Bright source peeling}
The removal of bright sources in Section \ref{subsub:peeling} to minimize the sidelobe effects is a computationally intensive process. In order to assess its efficiency, we made 2-s images at the moment of the burst appearance without peeling any sources. At 1.09 GHz where the bright sources are the most dominant due to the steep slope of their spectral index, it became practically unfeasible to differentiate genuine astronomical sources, such as the compact source, from sidelobes in the forms of horizontal stripes (see Figure \ref{5min_peel}), which could increase false detection rate. Furthermore, we also observed a decrease of $~ 10\%$ in S/N of the bursts at 1.48 GHz without applying the peeling.     

\subsubsection{Weighting scheme}
Further testing of the weight values during imaging shows that the bursts remain detected above our detection threshold mainly from natural to the 0.8 value of the Briggs robust parameter \citep{briggs1995high}. We decided to use robust 1 because it provides more resolution gain without much degradation in S/N from natural weighting mode. Given that a considerable number of MeerKAT dishes ($\sim 40 \; \text{antennas}$) is located around the $\sim$ 1 km inner core, briggs values that come close to uniformly weighted scheme deteriorate drastically the sensitivity due to the sparse distribution of the 2 s $(u,v)$ coverage.      

\subsubsection{Integration time} \label{subsec:integration}
\begin{figure}
	\centering
	
	\includegraphics[width=0.5\textwidth, height = 0.5\textwidth]{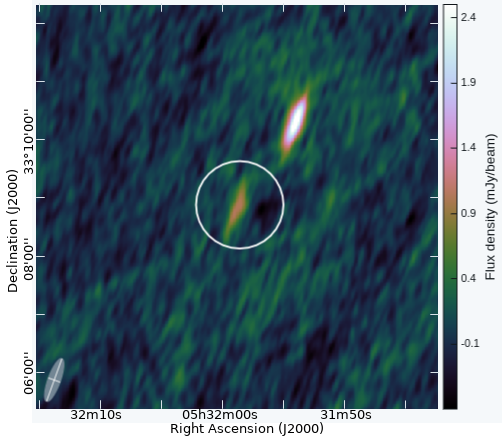}
	
	\caption{MeerKAT image of B11 using 8s integration time. The rms in this image is 150 $\mu$Jy.}   
	\label{fig:b11_8s}
\end{figure}

The majority of MeerKAT visibility measurement sets are dumped at 8 s integration time. In order to understand the detectability of similar bursts and fast transients in the MeerKAT archival data, we performed 8-s imaging using the same parameters as in Table \ref{imaging_param}. The overall resulting rms image is $150 \;\mu$Jy, yelding a fluence detection limit of $1.2$ Jy ms using the equation \ref{eq:frb_limit_fluence}. In this case, B07 and B11 are the bursts that remain above the detection limit but we only detected B11 with a reduced peak S/N ($\sim 6.8$) compared to the 2-s image values. Figure \ref{fig:b11_8s} displays the detection of B11 in the 8-s image. These findings show the importance of fast imaging and we recommend future MeerKAT observations of FRBs to be operated using the 2 second dump rate. The computational expenses of the short timescale imaging is discussed in Section \ref{subsec:compute}.

\subsubsection{Bandwidth} \label{subsec:bandwidth}
Instead of using the full band, we decided to split the data into two spectral windows to avoid similar averaging issue as discussed in Section \ref{subsec:integration}. Indeed, since the bursts does not appear in all the band, their signals are expected to be averaged out with larger bandwidths. Nonetheless, we produced 2-s images at the arrival times of the 11 bursts by combining all the available frequencies. As a results, we did not observe improvement of the image quality (rms $\sim 260 \mu$Jy ), and only B02, B03 and B11 were detected.  

\subsection{Image subtraction} \label{subsec:subtraction}
\subsubsection{Reference image} \label{subsec:refimage}
Since we already know the location of our target source and the arrival times of the bursts, it was straightforward to make the search towards a specific region of the sky. However, in a given survey of an unknown field, it is often challenging to identify a real transient source, especially at short integration time, where imaging artefacts can occasionally appear and increases the number of potential candidates. One of the most well known techniques used in transient search surveys is difference image analysis (DIA) \citep{tomaney1996expanding,bond2001real}, in which a reference image which is a model of the sky containing all steady-state sources, is subtracted from a new image. This section is not intended to derive a generalized image subtraction method for MeerKAT data, which can be a complex topic that could be explored in future studies, but rather a tentative framework to show its capability. Advanced discussions of the subject from previous works could be found in \citet{sanchez2019machine}, \citet{zackay2016proper}, \cite{bramich2008new} or \citet{alard2000image}.

In contrast to existing methods, where the reference image is constructed only from the preceding images of a sequence, we propose a new method where the subsequent images are also considered. Such procedure is inspired by the moving average technique widely known in economics \citep{choustatistical}. Given an image sequence of $N$ samples $\{I\} = \{I_1,I_2,I_3..,I_N\}$, the reference image $R_i$ of the $i-$th image, $I_i$, to be subtracted is defined as:

\begin{equation}
        \centering
        R_i = \frac{1}{2n} \sum_{\substack{j=i-n \\ j \neq i}}^{i+n} I_j , 
        \label{eq:subtraction}
\end{equation}

\noindent where $n$ is a tunable parameter that indicates the number of images in both sides of the $i-$th image to include in the summation. We tested our method with different values of $n$ and observed that the overall bursts S/N in the subtracted images increase by $20\%$ typically for $n = 1$ to $5$ and tend to follow a flat curve afterwards. Figure \ref{eq:subtraction} illustrates the evolution of S/N for B11 in the generated difference images for $n = 1$ to $n = 20$. For B11 only, we noticed a brief up and down trend after $n = 5$ that was not seen in the S/N curve of the other bursts. Based on Figure \ref{eq:subtraction}, it is tempting to claim that $n = 7$, which provided the maximum S/N is the optimal value, but we verified from manual inspection that it is simply due to small decrease of thermal noise from the added images at that period. Moreover, keeping the value of $n$ as small as possible is essential to preserve images with similar PSF that are produced from visibility with nearly the same $\textit{(u,v)}$ sampling distribution.                   

Figure $\ref{image_subtracted}$ demonstrates our difference imaging procedures and its effect in a $38'\times38'$ region around the burst position. The mean rms values of the subtracted images are about $3$ to $5\%$ lower than the original images and the bursts were still observed above $5\sigma$ except for B03 with $3\sigma$ level. The non detection of B03 is due to the presence of B02 when constructing the reference image at that moment, given that the appearance of the two bursts are only separated by 2 s. This shows that our proposed method is mainly limited to one-off events like most FRBs, or to repeating sources that emit pulses within sufficiently large time interval ($\geqslant 20$ s for $n = 5$). Additionally, constant sources in the $38' \times 38'$ area were almost completely removed and the occasional low level remaining artefacts are unclean subtraction from extended or bright sources. We show in Figure \ref{image_subtracted} the subtracted images for all 6 bursts. 

\begin{figure}
	\centering
	
	\includegraphics[width=0.5\textwidth, height = 0.4\textwidth]{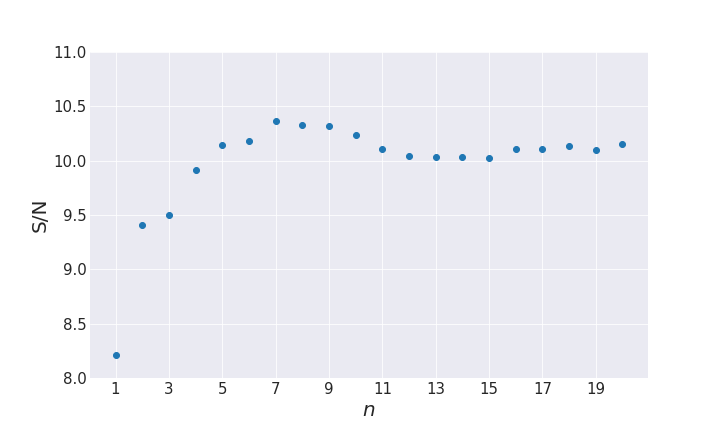}
	
	\caption{Signal-to-noise ratio (S/N) of B11 in the subtracted images for different values of $n$ (see equation \ref{eq:subtraction}), to build the reference image.}   
	\label{fig:B11_snr}
\end{figure} 

\begin{figure*}
	
    \centering
    \begin{subfigure}[b]{0.45\textwidth}
    	\includegraphics[width=\textwidth]{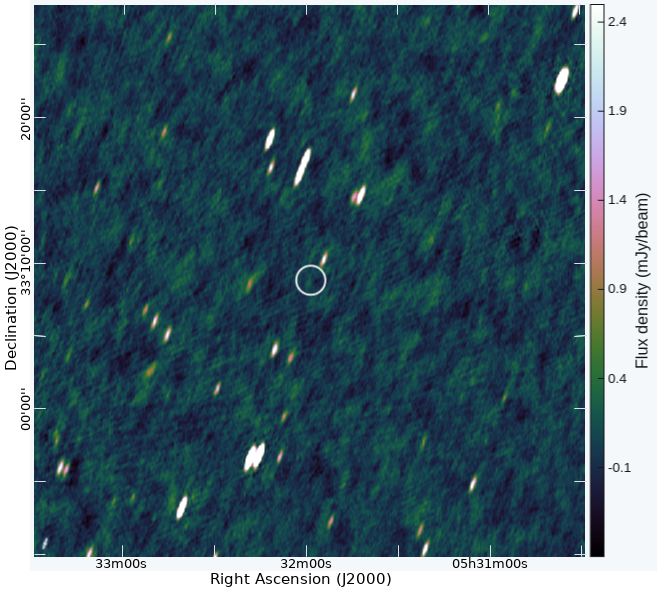}
    	\caption{}
    \end{subfigure}
    \centering
    \begin{subfigure}[b]{0.45\textwidth}
    	\includegraphics[width=\textwidth]{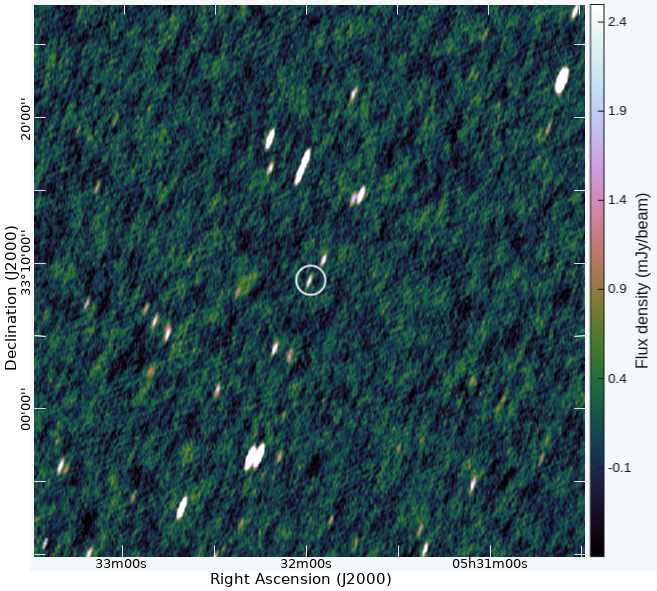}
    	\caption{}
    \end{subfigure}
    \centering
    \begin{subfigure}[b]{0.45\textwidth}
    	\includegraphics[width=\textwidth]{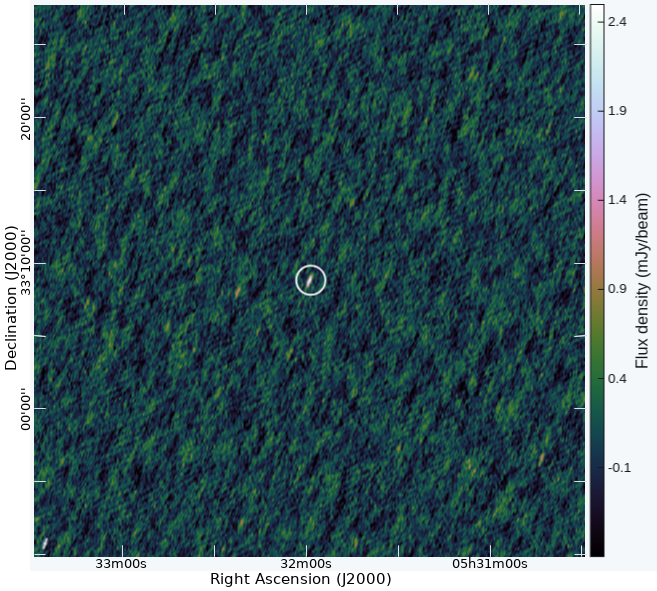}
    	\caption{}
    \end{subfigure}
    \centering
 
    \caption{ Illustration of an image subtraction procedure as described in Section \ref{subsec:subtraction}  for an area of $38' \times 38'$ around FRB121102. The figures are showing the reference image (a), B11 field before (b) and after (c) subtraction of the reference image. The subtracted image shows that most sources except B11 (White circle in the center) were removed.}
    \label{image_subtractedillus}
\end{figure*}

\begin{figure*}
	
    \centering
    \begin{subfigure}[b]{0.46\textwidth}
    	\includegraphics[width=\textwidth]{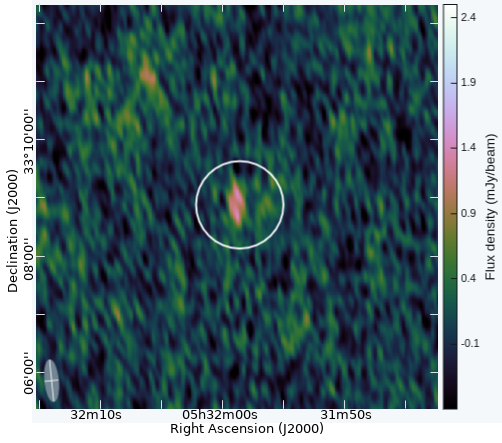}
    	\caption{Burst 02}
    \end{subfigure}
    \centering
    \begin{subfigure}[b]{0.46\textwidth}
    	\includegraphics[width=\textwidth]{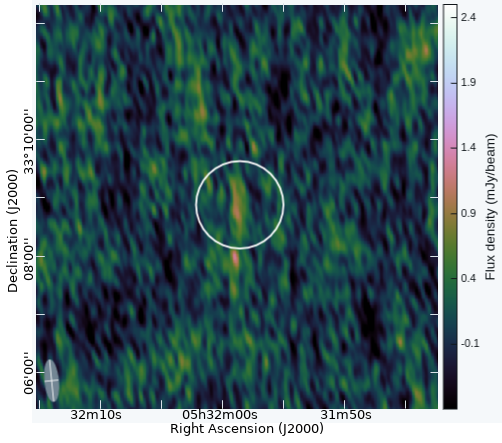}
    	\caption{Burst 03}
    \end{subfigure}
    \centering
    \begin{subfigure}[b]{0.46\textwidth}
    	\includegraphics[width=\textwidth]{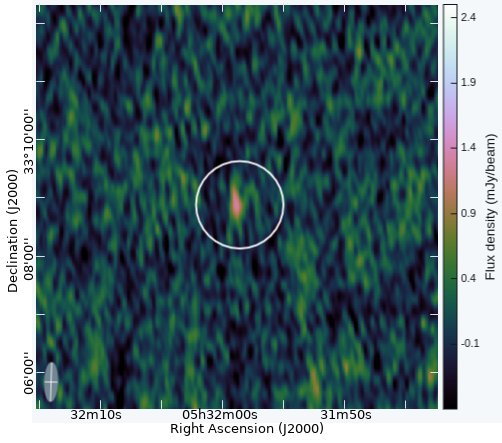}
    	\caption{Burst 05}
    \end{subfigure}
    \centering
    \begin{subfigure}[b]{0.46\textwidth}
    	\includegraphics[width=\textwidth]{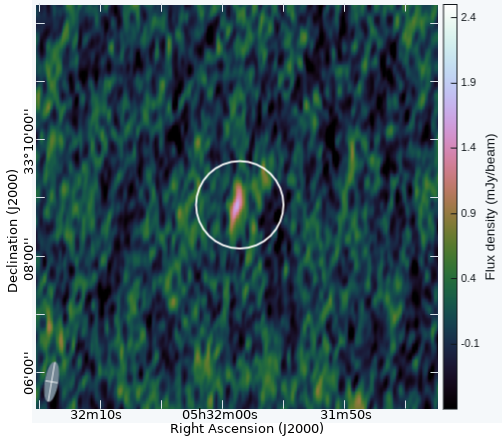}
    	\caption{Burst 07}
    \end{subfigure}

    \centering
    \begin{subfigure}[b]{0.46\textwidth}
    	\includegraphics[width=\textwidth]{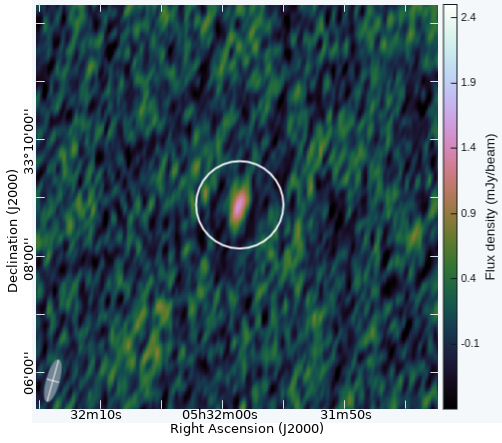}
    	\caption{Burst 08}
    \end{subfigure}
    \centering
    \begin{subfigure}[b]{0.46\textwidth}
    	\includegraphics[width=\textwidth]{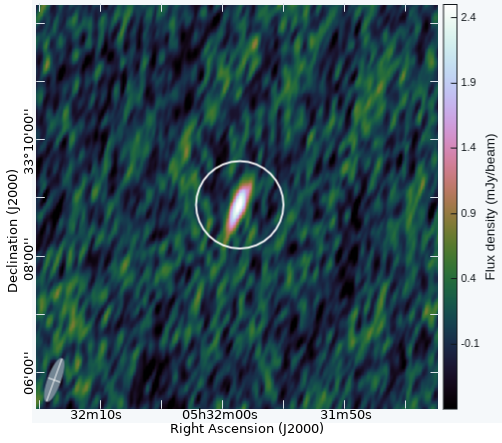}
    	\caption{Burst 11}
    \end{subfigure}
    \caption{MeerKAT 2-s image around an area of $7' \times 7'$ of the detected bursts after applying image subtraction (see Section \ref{subsec:subtraction}).}
    \label{image_subtracted}
\end{figure*}

\subsubsection{Blind search mode}
In order to show the potential advantages of our method in a blind search mode, we apply the procedure described in Section \ref{subsec:refimage} to all 2-s images produced from each 5 minute data and use the automated source finder PySE\footnote{\url{https://tkp.readthedocs.io/en/release4.0/tools/pyse.html?highlight=pyse}} from the TraP pipeline \citep{swinbank2015lofar} to identify the remaining point sources in the subtracted images. We set the detection threshold to 6$\sigma$ to decrease the number of candidates and mask the bright source positions to exclude eventual residual effects. We constrained the search to half of the images (about $1.25{^\circ}$ around the center) where 80$\%$ of the total flux density of the image resides and where the PSF is similar. As a results, only B07,B08 and B11 were detected and there were 23 false detections observed in a total of 1380 images, from which we calculated the fraction of false detection per synthesized beam element to be around $1.34\times10^{-6}$. In a blind search mode, automatic searches through the difference images are often performed with the use of machine learning techniques to classify real transients from artefacts \citep{goldstein2015automated,wright2015machine}, but in our case, the remaining candidates can also be reduced by searching for dispersed pulses in the dynamic spectrum of the raw voltage. 
Furthermore, more than $60\%$ of FRBs detected and published to date \citep{spanakis2022frbstats} have a fluence comparable or even stronger ($\sim$ 4 Jy ms to 3500 Jy ms) than B11 and we would expect to detect them with our method.

\subsubsection{Computational expenses}\label{subsec:compute}
Since our technique requires to perform imaging for each integration, it is important to assess the computational cost to run this approach. In terms of data storage, we are only considering to keep the 2-s dump rate around the time intervals where serious candidates are triggered and the data could be averaged to the standard integration time of 8 second otherwise. The data rate of the 2-s visibilities is 0.55 TB per hour for MeerKAT\footnote{\url{https://skaafrica.atlassian.net/wiki/spaces/ESDKB/pages/277315585/MeerKAT+specifications\#Visibility-integration-times}}, hence if we only keep for instance a 1 minute dataset due to an eventual candidates, only 9 GB of memory is needed.

With regards to imaging, the gridding step, which is the main operation that requires heavy processing, is often implemented with the use of a convolution kernel of $5 \times 5$ pixel wide, which means that 25 additions and multiplications will be operated for each visibility. One complex multiplication is a 7 float operations, whereas the addition involves 2 float operations, from which we can estimate that the convolution of one visibility requires $9 \times 50$ float operations. The number of visibility points involved in a stokes I 2-s imaging for the full MeerKAT array is given by \textrm{$n_{polarization} \times n_{channels} \times {n_{baseline}} = 2\times4096 \times 2016$}, thus the computation of gridding need about $\sim 7.5$ GB float operations, which could be achieved in less than 1 second with a GPU or a multicore CPU implementation \citep{veenboer2020radio}. 

\section{Conclusions}\label{sec:conclusion}

The detection of 11 bursts from FRB121102 was reported by \citet{caleb2020simultaneous} with the MeerKAT radio telescope using high time and frequency resolution filterbank data. In this work, we investigated the ability of MeerKAT to detect these bursts in 2-s snapshot images produced with visibility data.   

We detected 6 out of the 11 bursts in the images above a detection threshold of $5\sigma$ at 1.48 GHz. These represent the first detections of FRBs and radio transients in MeerKAT image produced with 2 seconds timescale. The 6 bursts were detected in accordance to our expectation with a fluence detection limit of $\sim 0.512$ Jy ms at the corresponding frequency. Additionally, the analysis of their properties revealed an $\sim$ arcsec precision of their localization in the images which is highly required to tie the FRB position with their potential host galaxy.  

We estimated from further investigation that the detection is strongly limited by the integration time and we recommend future MeerKAT observations of FRBs to be operated using the 2 second dump rate, which is the fastest supported integration period for this telescope.

We explored a new approach to the difference imaging analysis from which each image to be subtracted has their own reference image generated from their respective neighbor images. Such method takes advantage of the similarity of the $(u,v)$ distribution in the 2-s visibility data and could be adequately suitable to detect FRB-like transients in fast images.  

\section*{Acknowledgements}

This paper employs a MeerKAT data resulted from a Director’s Discretionary Time (DDT) proposal (Project ID: DDT-20190905-MC-01). The MeerKAT telescope is operated by the South African Radio Astronomy Observatory, which is a facility
of the National Research Foundation, an agency of the Department of Science and Innovation (DSI). The authors acknowledge the contribution of all those who designed and built the MeerKAT instrument.

Research reported in this paper is supported by the Newton Fund project, DARA (Development in Africa with Radio Astronomy), and awarded by the UK’s Science and Technology Facilities Council (STFC)
- grant reference ST/R001103/1. 

J.C.A. acknowledges the MeerTRAP team in collaboration with the ThunderKAT programme for allowing the access to the data in achieving this work. MeerTRAP is a project to continuously use the MeerKAT radio telescope to search the radio sky for pulsars and fast radio transients and to rapidly and accurately locate them. ThunderKAT is the MeerKAT Large Survey Projects (LSPs) for image-domain (explosive) radio transients.

M.C. acknowledges support of an Australian Research Council Discovery Early Career Research Award (project number DE220100819) funded by the Australian Government and the Australian Research Council Centre of Excellence for All Sky Astrophysics in 3 Dimensions (ASTRO 3D), through project number CE170100013.

M.C, F.J and B.W.S acknowledge funding from the European Research Council (ERC) under the European Union’s Horizon 2020 research and innovation pro- gramme (grant agreement No 694745).

We acknowledge the use of the ilifu cloud computing facility - www.ilifu.ac.za, a partnership between the University of Cape Town, the University of the Western Cape, the University of Stellenbosch, Sol Plaatje University, the Cape Peninsula University of Technology and the South African Radio Astronomy Observatory. The ilifu facility is supported by contributions from the Inter-University Institute for Data Intensive Astronomy (IDIA - a partnership between the University of Cape Town, the University of Pretoria and the University of the Western Cape), the Computational Biology division at UCT and the Data Intensive Research Initiative of South Africa (DIRISA).

\section*{Data Availability}
The data underlying this article will be shared on reasonable request to the corresponding authors.


\bibliographystyle{mnras}
\bibliography{main} 





\end{document}